\documentclass{qam-l}

\usepackage{amssymb}

\usepackage{graphicx}

\usepackage{slashbox}
\usepackage[numbers,sort&compress]{natbib}
\usepackage{color}

\copyrightinfo{2016}{Brown University}

\theoremstyle{definition}

\theoremstyle{remark}

\numberwithin{equation}{section}

\def\dd{\mbox{d}}
\newcommand{\vertcell}[2][c]{%
   \begin{tabular}[#1]{@{}c@{}}#2\end{tabular}}

\begin{document}

\title[Bistable radicalization/sectarian conflict model]{A bistable belief dynamics model for radicalization within sectarian conflict}


\author{Yao-Li Chuang}
\address{Dept. of Mathematics, CalState-Northridge, Northridge, CA 91330 \\
Dept. of Biomathematics, UCLA, Los Angeles, CA 90095-1766}

\author{Maria R. D'Orsogna}
\address{Dept. of Mathematics, CalState-Northridge, Northridge, CA 91330 \\
Dept. of Biomathematics, UCLA, Los Angeles, CA 90095-1766}
\email{dorsogna@csun.edu}

\author{Tom Chou}
\address{Departments of Biomathematics and Mathematics, UCLA, 
Los Angeles, CA 90095-1766}

\subjclass[2000]{Primary }

\date{\today}

\dedicatory{This work was made possible by support from grants ARO
  W1911NF-14-1-0472, ARO MURI W1911NF-11-10332, and NSF DMS-1516675.}

\begin{abstract}
\noindent
We introduce a two-variable model to describe spatial polarization,
radicalization, and conflict.  Individuals in the model harbor a
continuous belief variable as well as a discrete radicalization level
expressing their tolerance to neighbors with different beliefs.  A
novel feature of our model is that it incorporates a bistable
radicalization process to address memory-dependent social behavior.
We demonstrate how bistable radicalization may explain contradicting
observations regarding whether social segregation exacerbates or
alleviates conflicts.  We also extend our model by introducing a
mechanism to include institutional influence, such as propaganda or
education, and examine its effectiveness.  In some parameter regimes,
institutional influence may suppress the progression of radicalization
and allow a population to achieve social conformity over time. In
other cases, institutional intervention may exacerbate the spread of
radicalization through a population of mixed beliefs.  In such
instances, our analysis implies that social segregation may be a
viable option against sectarian conflict.
\end{abstract}

\maketitle


\section{Introduction}

\noindent
Recent years have seen the resurgence of ethnic, religious and racial
tension that have created rifts among communities once at peace.  In
many cases, friction has escalated towards violent conflict, ethnic
cleansing and at times even full-fledged civil wars that have
destabilized entire social and political systems
\cite{KAU96,SAM00,MAC01,KUN02,BAR04,CHA07,WEI13,BHA14}.
The development of viable intervention strategies to mitigate
radicalization and violence requires a thorough understanding of the
mechanisms underlying sectarian conflict.  Identifying the basic
ingredients that lead to the emergence of full scale conflict is
hindered by the complex nature of human behavior. Instead of
simplistic, universal interpretations and solutions, one is often left
with contradictory observations and outcomes.

In particular, there is controversy as to whether social segregation
should be employed to manage sectarian conflicts
\cite{DIX05,BHA14}. Some studies suggest that inter-ethnic or
inter-communal contacts raise tension and that it is beneficial to
keep rival communities separate until tensions dissipate
\cite{OLE93,KAU96,SAM00,MAC01,BAR04,CHA07,WEI13}. Others have
concluded that ethnically mixed environments encourage inter-ethnic
friendship and trust, while segregation leads to prejudice and
antagonistic behavior \cite{KUN02,BAR04,DIX05,RYD13,BHA14}.
These contradicting conclusions reveal the context-dependent nature of
human social behavior. 

One of the goals of this paper is to present a mathematical framework
that may help resolve basic observations of belief dynamics,
radicalization, and conflict.  Social studies have shown that humans
often hysteretically switch behaviors, perceiving and reacting to the
same information in drastically different ways because of different
past experiences and circumstantial contexts.  This hysteretic
switching behavior also applies to general tolerance towards others
and their views.  Similar socio-economic environments in some cases
have led to peaceful coexistence between communities, in others to
conflict. Within the context of radicalization we model this
hysteresis using a memory-dependent or ``bistable'' response to the
social environment \cite{LAW09,LAW10}.  To quantify this
mode-switching behavior, we draw inspiration from the physical
sciences, where bistability is ubiquitous; for example, in
ferromagnetism where materials switch their magnetic alignment as an
external field is varied \cite{BOZ51}.

\begin{figure}
  \begin{center}
      \includegraphics[width=4.5in]{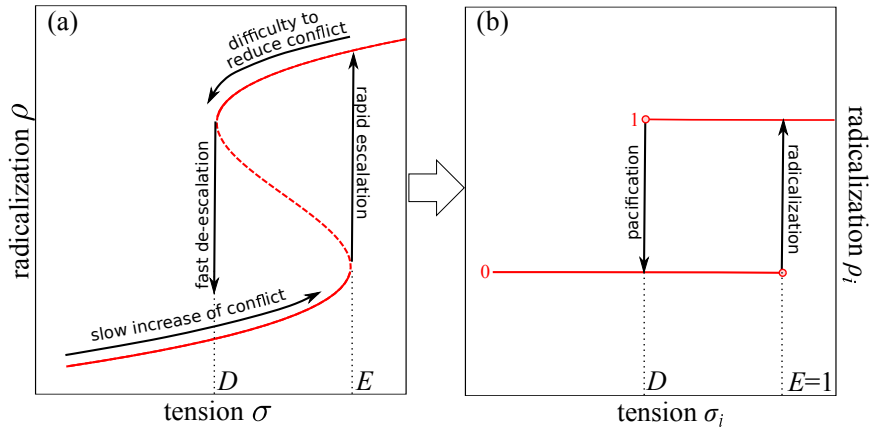}
\end{center}
  \caption{(a) A typical bistability curve. The curve qualitatively
    captures the essence of escalating conflicts and radicalization
    $\rho$ due to rising tension $\sigma$. Bifurcation values of the
    tension $\sigma$ are labelled $D$ and $E$.  (b) A
    simplified piecewise constant description of the bistable
    dependence of $\rho_i$ on $\sigma_i$.
             \label{FIG:BISTABLE} }
\end{figure}

Fig.~\ref{FIG:BISTABLE}(a) depicts the hysteresis curve of a system in
which a bistable state variable $\rho$ ($y$-axis) is driven by an
independent regulating variable $\sigma$ ($x$-axis).  The curve
represents the equilibrium solution to, \textit{e.g.}, a differential
equation for $\rho$ in which $\sigma$ is a controlling parameter.  The
solid parts of the curve indicate stable values of $\rho$, while the
dashed segment are unstable solutions.  The functional dependence of
$\rho$ on $\sigma$ yields a window of values $D < \sigma < E$ in which
two stable solutions can arise.

Within the context of belief/radicalization dynamics, $\rho$ may
represent the degree of radicalization of a population or an
individual that is driven by social tension $\sigma$.  An interesting
and frequently observed phenomenon is that of a slowly deteriorating
political, economic or social situation (increasing $\sigma$) which
seems under control but abruptly escalates.  The lower solid curve
indicates a less radicalized population (low $\rho$) that favors
peaceful coexistence with others of different views. Increasing social
tension can force $\rho$ to transition from the lower to the upper
solid curve at $\sigma \ge E$. The upper curve represents a highly
radicalized population that is non-tolerant towards those with
different views. The ``bifurcation point'' $E$ thus marks a sudden
escalation of sectarian conflict which can be triggered by random
events. Once the situation escalates, it is often very difficult to
restore peace, as $\rho$ remains on the upper solid curve even if
$\sigma$ is decreased back below $E$.  Peace can only be restored if
sufficient effort is made to further reduce $\sigma$ below the lower
bifurcation point $D < E$.  The hysteresis between high and low
radicalization levels may help shed light on contradicting reports
regarding whether the promotion of ethnic mixing or segregation is the
best approach to achieve a state of peaceful coexistence.  Just like
tension can rapidly escalate, it may also rapidly de-escalate.  One
example might be the decades-long Northern Ireland conflict.  As late
as 1993, some scholars were still very pessimistic on a possible
peaceful resolution of the Catholic/Protestant conflict, stating that:
``the cruel conflict will continue, apparently with no end in
sight...''  \cite{OLE93,DIX05}. However the 1994 IRA ceasefire quickly
lead to the 1998 signing of the Good Friday Agreement, marking the end
of ``the Troubles.''  To phenomenologically incorporate bistability
one can, without loss of generality, adopt a simplified description of
the relationship between radicalization and tension, as shown in
Fig.~\ref{FIG:BISTABLE}(b).

%

To understand influence of behavioral memory on the spread of
radicalization and conflict, we will incorporate this bistable
dynamics into the belief dynamics model of DeGroot, which describes an
individual's the belief as a one-dimensional continuous variable
bounded by two extreme limits \cite{DEG74,FRI99,GOL10}. Over time,
individuals may change their opinions by interacting with others.  In
the DeGroot model, originally introduced to study the formation of
consensus opinions in a network, conformity is the only ensured
outcome.  The inability to form heterogeneous distributions of
opinions, or ``persistent disagreement'', limits the applicability of
the model to ethnically or ideologically divergent societies
\cite{ACE11,ACE13,YIL13,EGE13}. Extensions of the DeGroot model have
been proposed to induce disagreement, such as the popular ``bounded
confidence'' model, where individuals interact only with those holding
similar opinions, defined by an opinion range called the bounded
confidence \cite{DEF00,KRA00,WEI02,FEL13,FEL14}.

Another approach is taken via ``opinion opposition'' models that
introduce agents of ``nonconformity'' who adopt contrarian views and
cause polarized beliefs and disrupt the formation of consensus
\cite{EGE13,JAV14}. Such models share similarities with spin glass
Ising models that describe a mixture of ferromagnetic and
antiferromagnetic molecules; the former tend to align their spins with
neighbors, while the latter anti-align.  Since sectarian conflicts
usually arise and progress through direct conflict of opinion, we will
adopt an opinion opposition approach rather than a bounded confidence
approach.  Assuming that opinion differences among individuals causes
tension, we incorporate bistability to quantify the level of
radicalization that may prompt an individual to radicalize.

We note that relatively peaceful, albeit tense, coexistence between
communities of different backgrounds can be ensured by a strong or
influential player, such as the state, a dictator, inter-communal
institutions, or the international community. The removal of such a
player correlates with outbreaks of violent conflicts
\cite{DIX05,HAG13,WEI13,CAR14}. Thus, we will also incorporate the
influence of a central figure, modeled as a globally connected player
exerting institutional influence similar to the concept of media
influence on locally connected networks \cite{THR12}.

In the next Section, we present the details of our basic model of
radicalization and sectarian conflict. We then augment the basic model
to include government propaganda and explore how it influences
sectarian conflict. One of our aims is to use our model to inform
strategies that can stop radicalization and sectarian violence from
spreading among an ethnically mixed population without employing
population segregation as a peace-keeping method.  Results of our
analysis will be presented in the Results and Discussion Section,
where parameter dependence will also be explored.

\section{Bistable lattice model of conflict}

\begin{figure}
  \begin{center}
      \includegraphics[width=3.3in]{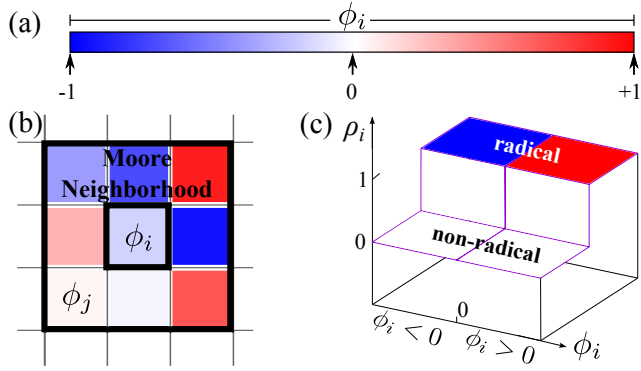}
\end{center}
  \caption{(a) Schematic of the belief variable $-1 \le \phi_i \le
    1$. Positive and negative values respectively represent the
    degrees of belief towards the extreme red and the blue
    ideologies. The color-coding indicates different values of
    $\phi_i$, with darker colors designated for more zealous
    beliefs. (b) Definition of nearest neighbors $[i]$. We define
    nearest neighbors of agent $i$ according to its Moore
    neighborhood, which include the eight lattice sites surrounding
    site $i$. (c) Definition and color-code of radicalization
    $\rho_i$. The discrete value of $\rho_i$ is determined via the
    simplified piecewise constant function in
    Fig.~\ref{FIG:BISTABLE}(b). $\rho_i = 1$ represents radicals and
    is colored red or blue, depending on the corresponding sign of
    $\phi_i$.  Non-radicals have $\rho_i = 0$ and are colored white,
    regardless of their belief value $\phi_{i}$.
            \label{FIG:SCHEMATIC} }
\end{figure}

We assume a two-dimensional $N \times N$ site lattice model where each
site $i$ is populated by an agent. Two dynamical variables are
associated with each agent: a continuous ``belief'' variable $-1 \leq
\phi_i (t) \leq 1$ indicating the strength of belief in an ideology of
agent $i$, and a discrete ``radicalization'' variable $\rho_i (t) \in
\{0,1\}$ indicating the intolerance of agent $i$ towards a very
different ideology.  Since radicalization usually leads to conflict,
we will use the two concepts interchangeably. Radicals cause conflict;
non-radicals allow for peaceful coexistence.  As shown in
Fig.~\ref{FIG:SCHEMATIC}(a), we color-code the two extreme belief
values $\phi = -1$ and $\phi = +1$ blue and red respectively, while
lighter colors indicate intermediate values. Despite the continuous
belief values, we divide the population into red ($\phi_i>0$) and blue
($\phi_i<0$) groups and refer them as two sects.  We assume a fully
occupied periodic lattice without empty sites, and that the occupying
agents do not migrate.

The values $\phi_i (t)$ and $\rho_i (t)$ evolve over time via
nearest-neighbor interactions. 
Nearest neighbors are defined using the ``Moore neighborhood,'' where
eight grid sites surrounding site $i$ are considered, as shown in
Fig.~\ref{FIG:SCHEMATIC}(b). In the following subsections we describe
the model that governs the evolution of $\phi_{i}(t)$ and
$\rho_{i}(t)$.

\subsection{Belief and radicalization}

The magnitude of belief $\vert \phi_i \vert$ measures the level of
enthusiasm of agent $i$.  Individuals with $\vert \phi_i \vert \approx
1$ are belief zealots while those with $\vert \phi_i \vert \approx 0$
are belief apathetics.  In addition to the belief variable $\phi_{i}$,
a discrete radicalization variable $\rho_i\in \{0,1\}$ describes how
an agent perceives other beliefs.  An intolerant individual at site
$i$ will be assigned $\rho_{i} = 1$ and referred to as a
radical. Conversely, a tolerant non-radical will be described by
$\rho_i = 0$.  Within the context of our model, $\phi$ and $\rho$
describe distinct attributes.  Zealots can be tolerant of the opposite
sect and be non-radical.  For example, zealots may be deeply religious
individuals who at the same time are accepting of others' beliefs.

The site-specific variables $\phi_{i}$ and $\rho_{i}$ depend on each
other through an intervening social tension variable $\sigma_{i}$.
The basic mechanism for this interplay is that the tension
$\sigma_{i}$ felt by agent $i$ arises from differences in belief
($\phi_{i}-\phi_{j}$) between agents $i$ and $j$.  In turn, the level
of tension $\sigma_{i}$ determines the radicalization state $\rho_{i}$
of agent $i$, who finally adjusts its belief $\phi_i$ accordingly. As
mentioned in the Introduction, we will assume $\rho_i$ to be bistable.
The dynamical model is mathematically described below.
%
%

\subsection{Tension and belief adaption}

The tension $\sigma_{i}$ perceived by agent $i$ is determined as follows:

\begin{equation}
   \sigma_i [\rho(t), \phi(t)] \equiv \sum_{j \in [i]}
       J(\rho_i, \rho_j) \left( \phi_i - \phi_j \right)^2,
   \label{EQ:TENSION}
\end{equation}

\noindent
where the coupling coefficient $J(\rho_i, \rho_j)$ characterizes
the sensitivity of agent $i$ with radicalization $\rho_i$ towards
the view expressed by agent $j$ with radicalization
$\rho_j$. The sum over $j$ is then taken
over Moore neighborhood of $i$, $[i]$, as shown in 
Fig.~\ref{FIG:SCHEMATIC}(b).

Eq.~\ref{EQ:TENSION} allows for tensions to increase when neighbors
$i$ and $j$ have different belief levels as modulated by $J(\rho_i,
\rho_j)$. By construction, if all sites neighboring $i$ carry the same
belief value $\phi_{i}$, the perceived tension $\sigma_i = 0$.  The
functional dependence of $J(\rho_i, \rho_j)$ will be defined in the
Model Parameters section.  Since the maximum of $\vert \phi_i - \phi_j
\vert = 2$, $0 \le \sigma_i \le 32\, {\rm max}(J)$, where ${\rm
  max}(J)$ is the maximum of $J(\rho_i, \rho_j)$.

The discrete value assigned to $\rho_i$ is determined by the
piecewise-constant hysteresis function illustrated in Fig.~\ref{FIG:BISTABLE}(b)
and depends on whether $\sigma_i$, determined in
  Eq.~\ref{EQ:TENSION}, exceeds a ``radicalization point'' $E$, is
below a ``pacification point'' $D$, or lies in between.  
The bistable dependence of $\rho_i$ on $\sigma_i$ can be
expressed as follows
\begin{equation}
\rho_i (\sigma_i (t)) \left\{ \begin{array}{ll}
         = 1 & \textrm{ if } \sigma_i (t) > E, \\
         = 0 & \textrm{ if } \sigma_i (t) < D, \\
         \textrm{unchanged} & \textrm{ otherwise}.
   \end{array}\right.
   \label{EQ:RHO}
\end{equation}

\noindent
Since $D$ and $\sigma_i$ (and consequently $J$ in
Eq.~(\ref{EQ:TENSION})) can be rescaled by $E$; without loss of
generality, we can set $E \equiv 1$.  Phenomenologically,
Eq.~(\ref{EQ:RHO}) allows high tension to drive a non-radical toward
radicalization, while low tension may pacify a radical.

We assume the radicalization state $\rho_i$ feeds back to $\phi_i$ via
a modified continuous-time DeGroot model to include contrarian
behavior as follows.

\begin{equation}
   \frac{\dd \phi_i(t)}{\dd t} = 
          \sum_{j \in [i]} k(\rho_i, \rho_j, \phi_i, \phi_j) 
                           \left( \phi_j - \phi_i \right)
   \label{EQ:ZEAL_ADJUSTMENT}
\end{equation} 

\noindent
where $k(\rho_i, \rho_j, \phi_i, \phi_j)>0$ is the rate of change of
belief from $\phi_i$ towards $\phi_j$. Negative $k$ indicates a
$\phi_i$ that drifts away from $\phi_j$. The functional dependence of
$k$ will be defined in Model Parameters section

Note that Eq.~(\ref{EQ:ZEAL_ADJUSTMENT}) can also be written in the
form $\phi_i (t + \dd t) = \sum_{j} M_{ij} \phi_j (t)$, where $M_{ij}=
k(\rho_i, \rho_j, \phi_i, \phi_j) \dd t$ for $j\in [i]$, and $M_{ii}
\equiv (1 - \sum_{j \in [i]} k(\rho_i, \rho_j, \phi_i, \phi_j)) \dd
t$.  For discrete-time DeGroot models $\dd t = 1$ and the matrix
$\mathbf{M}$ is known as the ``trust matrix'' satisfying $\sum_{j}
M_{ij} = 1$.  To prevent $\phi_i (t)$ from exceeding the bounds, we
further implement no flux boundaries by requiring $k \to 0$ at $\phi_i
= \pm 1$.

The rules governing the belief value $\phi_{i}$, the intolerance level
$\rho_{i}$, and the perceived tension $\sigma_{i}$, are given by
Eqs.~\ref{EQ:ZEAL_ADJUSTMENT}, \ref{EQ:RHO}, and \ref{EQ:TENSION},
respectively. With initial conditions and definitions of the parameter
functions $J(\rho_i, \rho_j)$ and $k(\rho_i, \rho_j, \phi_i, \phi_j)$
in the next section, these equations fully define our bistable
radicalization and belief dynamics model.

\subsection{Model parameters}

In this subsection, we define the dependence of $J(\rho_i, \rho_j)$
and $k(\rho_i, \rho_j, \phi_i, \phi_j)$ and then determine the number
of independent parameters of the model.  We first discuss the coupling
function $J(\rho_i, \rho_j)$ and assume that interactions with or
between radicals heighten the sensitivity towards belief diversity,
resulting in higher social tension.  We thus assign

\begin{equation}
   J(\rho_i, \rho_j) = \left\{ \begin{array}{ll}
         J_- & \textrm{if } \rho_i = \rho_j = 0, \\
         J_+ & \textrm{otherwise},
   \end{array}\right.
   \label{EQ:J}
\end{equation}

\noindent
where $J_+ \ge J_- \ge 0$ quantify high and low sensitivities.

For the rate of change of belief presented in
Eq.~(\ref{EQ:ZEAL_ADJUSTMENT}), it is required that $k(\rho_i, \rho_j,
\phi_i, \phi_j) \to 0$ at $\vert \phi_i \vert = 1$ to prevent $\phi_i$
from exceeding the bounds. For $\vert \phi_i \vert < 1$, we set
$k(\rho_i, \rho_j, \phi_i, \phi_j) = \pm 1$ to most simply describe
conformation and dissension. If agent $i$ finds the belief of its
neighbor $j$ agreeable, $\phi_i$ ``ferromagnetically'' adjusts towards
$\phi_j$ at the rate $k(\rho_i, \rho_j, \phi_i, \phi_j) =
1$. Conversely, if neighbor $j$ antagonizes agent $i$, $k(\rho_i,
\rho_j, \phi_i, \phi_j) = - 1$ and $\phi_i$ shifts away from $\phi_j$,
resulting in an ``antiferromagnetic'' behavior. Finally
  we assume the following qualitatively reasonable rules to determine
whether belief conformation or dissension occurs.

\begin{enumerate}
   \item A non-radical ($\rho_{i}=0$) conforms to the beliefs of
     neighboring non-radicals but dissents from the beliefs of
     radicals ($\rho_j = 1$), regardless of their belief $\phi_{j}$ of
     the neighbors. In this case
     \begin{equation}
        k(\rho_i, \rho_j, \phi_i, \phi_j) = \left\{
         \begin{array}{rll}
             1  & \textrm{ if } \rho_i = 0 & \textrm{ and } \rho_j = 0 \\
            -1  & \textrm{ if } \rho_i = 0 & \textrm{ and } \rho_j = 1 
        \end{array}
        \right.
        \label{EQ:NONRADICAL_r}
     \end{equation}
   \item A radical conforms to the beliefs of neighbors of the same
     sect and dissents from the beliefs of neighbors of the opposite
     sect, regardless of their radicalization level. In this case
     \begin{equation}
        k(\rho_i, \rho_j, \phi_i, \phi_j) = \left\{
         \begin{array}{rll}
             1  & \textrm{ if } \rho_i = 1 
                & \textrm{ and } \phi_i \phi_j \ge 0 \\
            -1  & \textrm{ if } \rho_i = 1 
                & \textrm{ and } \phi_i \phi_j < 0
        \end{array}
        \right.
        \label{EQ:RADICAL_r}
     \end{equation}
\end{enumerate}

\begin{figure}
  \begin{center}
      \includegraphics[width=5in]{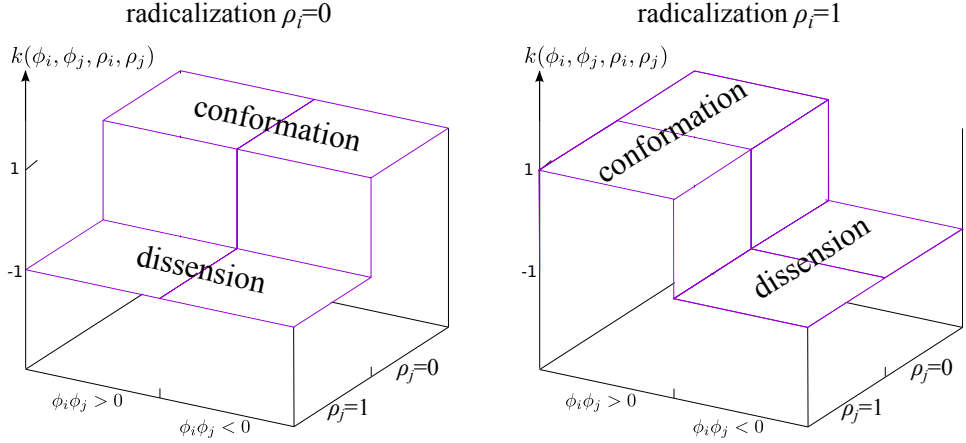}
\end{center}
  \caption{An illustration of the functional dependence of the belief
    evolution rate $k(\rho_i, \rho_j, \phi_i, \phi_j)$.  Depending on
    $\rho_i$, radicals and non-radicals adopt conformation ($k = 1$)
    and dissension ($k = -1$) behavior differently. Non-radicals
    ($\rho_i = 0$, left panel) determine $k$ based on $\rho_j$;
    radicals ($\rho_i = 1$, right panel) determine $k$ based on the
    sign of $\phi_i \phi_j$, \textit{i.e.}, whether individual $j$
    belongs to the same sect.
            \label{FIG:RATE} }
\end{figure}

\noindent
The above assignment of $k(\rho_i, \rho_j, \phi_i,\phi_j)$ is
summarized in Table~\ref{TABLE:RATE} below and is illustrated in
Fig.~\ref{FIG:RATE}. The discontinuity of $k(\rho_i, \rho_j,
\phi_i,\phi_j)$ at $\vert \phi_i \vert = 1$ can be made continuous by
setting $k(\rho_i, \rho_j, \phi_i,\phi_j) = \pm \left[ 1 - \tanh
  ((\vert \phi_i \vert - 1)/\epsilon)/2 \right]$ with an infinitesimal
parameter $\epsilon \ll 1$. For numerical simulations, we may choose
$\epsilon$ at the same order as the time step size.

%
\begin{table}[htb]
\begin{tabular}{|c||c|c|c|c|}
\hline
\backslashbox{$\phi_j $ \\ $\rho_j  $}
             {$\phi_i  $ \\ $\rho_i  $} & 
  \vertcell{$\phi_i  >0$\\$\rho_i  =1$} & 
  \vertcell{$\phi_i  >0$\\$\rho_i  =0$} &
  \vertcell{$\phi_i  <0$\\$\rho_i  =1$} & 
  \vertcell{$\phi_i  <0$\\$\rho_i  =0$} \\
\hline
\hline
 \vertcell{$\phi_j  >0$\\$\rho_j  =1$} & 
   $+1$ & $-1$ &
   $-1$ & $-1$ \\ 
\hline
 \vertcell{$\phi_j  >0$\\$\rho_j  =0$} & 
   $+1$ & $+1$ &
   $-1$ & $+1$ \\ 
\hline
 \vertcell{$\phi_j  <0$\\$\rho_j  =1$} & 
   $-1$ & $-1$ &
   $+1$ & $-1$ \\ 
\hline
 \vertcell{$\phi_j  <0$\\$\rho_j  =0$} & 
   $-1$ & $+1$ &
   $+1$ & $+1$ \\ 
\hline
\end{tabular}
\begin{center}
  \caption{The table lists the value of $k(\rho_i, \rho_j, \phi_i,
    \phi_j)$, depending on $\rho_i $ and $\rho_j $, as well as $\phi_i
    $ and $\phi_j $. }
\label{TABLE:RATE}
\end{center}
\end{table}
%
%
Note that $k(\rho_i, \rho_j, \phi_i, \phi_j)$ need not be symmetric
with respect to the interchange of $i$ and $j$ since individual $i$'s
reaction toward individual $j$ will in general be different from that
of $j$'s toward $i$.  This is a major difference between human
interactions and physical interactions, which are typically symmetric.

With the definition of $ J(\rho_i, \rho_j) = J_\pm$ and $k(\rho_i,
\rho_j, \phi_i, \phi_j) = \pm 1$, our equations now have three
independent constant parameters: $D$, $J_+$ and $J_-$.  Other
adjustable parameters not in the equations include the size $N$ of the
periodic lattice and the initial conditions. We vary the initial
red-to-blue population ratio, which we denote as $R(0)$, and unless
specified otherwise, we set $J_-=0.03$, $J_+=0.6$, $D=0.1$, $R(0)=3/7$
and $N=100$ as our default parameter values for simulations of
Eqs.~(\ref{EQ:TENSION})-(\ref{EQ:ZEAL_ADJUSTMENT}).  These values are
chosen based on our extensive parameter sweep as described in Results
and Discussion section.

\subsection{Institutional influence}

While the basic model
(Eqs.~(\ref{EQ:TENSION})-(\ref{EQ:ZEAL_ADJUSTMENT})) describes the
spread of radicalization, we may also wish to include intervention
strategies that may alleviate conflict.  Historically, a more peaceful
coexistence of divided populations are facilitated by the presence of
a strong or influential central figure, such as a state, a dictator,
inter-communal institutions, or the international community
\cite{DIX05,HAG13,WEI13,CAR14}.  While such a central figure can
influence various facets of a society, in this paper we mainly
consider how the outreach of government institutions affects social
tension.

%
\begin{figure}
  \begin{center}
      \includegraphics[width=3.3in]{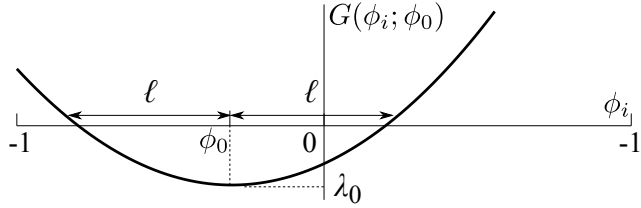}
\end{center}
  \caption{A global institutional influence function
    $G(\phi_i;\phi_{0})$. The function is defined by three parameters:
    the institutional stance of belief $\phi_0$, the strength of the
    influence $\lambda_0$, and the broadness of institutional messages
    $\ell$. For $\phi_0-\ell < \phi_i < \phi_0 + \ell$,
    $G(\phi_i;\phi_{0})<0$ and the tension $\sigma_i$
    decreases. However, outside this range, $\sigma_i$ increases.
            \label{FIG:LAMBDA} }
\end{figure}

We model a governmental institution as a globally connected player
that adopts a stance $\phi_0$ on the belief scale \cite{THR12}.
Institutional publicity or incentives may appease individuals holding
similar beliefs to $\phi_0$, causing a reduction of the social tension
they perceive. However, for individuals with significantly different
beliefs compared to $\phi_0$, the perceived tension may increase.  We
model the influence of the institutions on the social tension
perceived by agent $i$ via a simple three-parameter quadratic function
\begin{align}
  G ( \phi_i ) = & \frac{\lambda_0}{\ell^2} \left(
    \phi_i - \phi_0 \right)^2 - \lambda_0,
     \label{EQ:LAMBDA}
\end{align}
as plotted in Fig.~\ref{FIG:LAMBDA}. Under governmental influence, the
social tension obeys
\begin{align}
   \sigma_i [\rho(t); \phi(t)] = &
       \displaystyle \sum_{j \in [i]} J(\rho_i, \rho_j)
       \left( \phi_i - \phi_j \right)^2
       + G(\phi_i;\phi_{0}).
   \label{EQ:TENSION2}
\end{align}
The constant $\lambda_0$ represents the strength of the institutional
influence and is proportional to, say, the available resources and
invested efforts. The half distance $\ell$ between the two $x$-axis
intercepts characterises the broadness of the institutional
message. Within the range $\phi_0 - \ell < \phi_i < \phi_0 + \ell$,
$G(\phi_i)<0$. Here the institutional message is assumed to be
appeasing to individual $i$, leading to the reduction of its social
tension $\sigma_i$.  However, individuals with beliefs outside of this
range will experience an increased tension.

For simplicity, we assume that the institutional stance $\phi_0$ does
not directly sway a belief $\phi_i$, leaving Eqs.~(\ref{EQ:RHO}) and
(\ref{EQ:ZEAL_ADJUSTMENT}) intact. However, an institution may
indirectly steer the beliefs of a general population towards its
stance by reducing social tensions and thus encouraging conformity
towards $\phi_0$.

In the following section, we first identify the scenarios that lead to
the spread of radicalization in the basic model
Eqs.~(\ref{EQ:TENSION})-(\ref{EQ:ZEAL_ADJUSTMENT}) without
institutional influence. We then include such institutional influence
by replacing Eq.~(\ref{EQ:TENSION}) with Eq.~(\ref{EQ:TENSION2}) and
explore the outcomes.

\section{Results and discussion}

We first examine the basic model
Eqs.~(\ref{EQ:TENSION})-(\ref{EQ:ZEAL_ADJUSTMENT}) without
institutional influence to investigate the dependence of $\phi_{i}$
and $\rho_{i}$ on the five adjustable parameters: $J_+$, $J_-$, $D$,
$N$, and $R(0)$.  Simulations of the basic model are carried out by
numerically integrating
Eqs.~(\ref{EQ:TENSION})-(\ref{EQ:ZEAL_ADJUSTMENT}) using a
semi-implicit method to update the levels of belief $\phi_i $ and
radicalization $\rho_i$.  The numerical discretization is detailed in
the Appendix.

\begin{figure}[h]
  \begin{center}
      \includegraphics[width=3.3in]{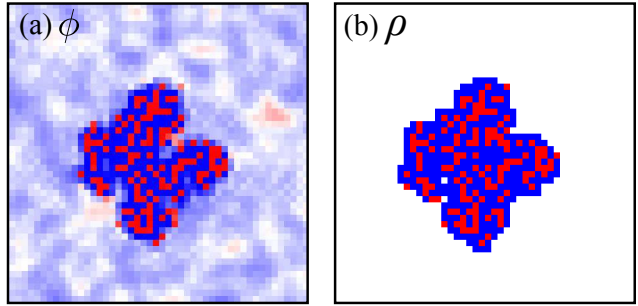}
\end{center}
  \caption{A snapshot of the spatial distribution of (a) $\phi_i$ and
    (b) $\rho_i$ during a simulation. (a) The color codes (see
    Fig.~\ref{FIG:SCHEMATIC}(a)) reflect the intensity level of
    belief, with red and blue indicating the two opposing opinions.
    (b) The corresponding radicalization values $\rho_{i}$ using the
    color codes in see Fig.~\ref{FIG:SCHEMATIC}(c). The simulation is
    initiated with $\rho_i = 0$ and randomly drawn $\phi_i$ from a
    uniform distribution such that $R(0)=3/7$. A radicalized agent
    with $\rho^* = 1$ and $\phi^* = 0.9$ is seeded at the center
    of the lattice, which triggers the spread of radicalized
    population. The simulation is conducted on a $100 \times 100$
    lattice, but the images here are cropped to better show the
    radicalized population at the center. The other parameters are set
    to the default values.
    \label{FIG:COMP}}
\end{figure}

The initial conditions of the simulation were set at $\rho_i=0$ and
randomly drawing values of $\phi_i$ from a uniform distribution.  We
further rebalanced $\phi_{i}$ such that the $\phi_{i}>0$ to
$\phi_{i}<0$ (red-to-blue) ratio was $R(0)=3/7$. Next, we placed a
radical agent ($\rho^* = 1$) with belief $\phi^* = 0.9$ at the center
of the lattice. For the rest of the paper, $\rho^*=1$ and $\phi^*=0.9$
will be used as the initial values of the radical agent at the center
of the lattice if such a seed is planted.  The results are
qualitatively similar for sufficiently extreme values of $\phi^*$
($\phi^* \gtrsim 0.9$ or $\phi^* \lesssim -0.9$). Uncertainty however
rises with smaller $\vert \phi^* \vert$ as the radical seed tends to
be increasingly pacified at the onset of our simulations.  A snapshot
of $\phi_i$ and $\rho_i$ at $t=1$ is shown in Figs.~\ref{FIG:COMP}(a)
and (b), respectively, and the default parameter values are used for
the simulation.  Note that by normalizing $\vert k(\rho_i, \rho_j,
\phi_i, \phi_j) \vert$ our simulation time $t$ is defined on the
belief-changing time scale.
%
%
In Fig.~\ref{FIG:COMP}(a) we use the color codes in
Fig.~\ref{FIG:SCHEMATIC}(a) to depict $\phi_i$ for each individual
$i$, with darker red/blue colors representing more extreme views among
the respective sects. In Fig.~\ref{FIG:COMP}(b) we plot the
corresponding $\rho_i$ using the color codes in
Fig.~\ref{FIG:SCHEMATIC}(c), where radicals are marked by red/blue
grids and non-radicals white. As can be seen, radicals tend to exhibit
a more extreme level of belief than non-radicals. The latter mostly
conform toward relatively neutral beliefs if not in contact with
radicals.  However, one can still see darker spots in
Fig.~\ref{FIG:COMP}(a) in the regions corresponding to non-radical
sites in Fig.~\ref{FIG:COMP}(b).  This shows that regions in which
zealots are not radicalized can be sustained, and peaceful coexistence
can be achieved. During the simulation, non-radicals can be
radicalized by their radical neighbors, leading to an outward spread
of radicalization from the initially planted radical seed.
Considering that radicalization often precedes conflicts, this
``contagion'' effect may be referred to as ``escalation diffusion'' of
conflicts, which was identified as a dominant mechanism driving the
spread of conflicts \cite{SCH11}.

\begin{figure}[h]
  \begin{center}
      \includegraphics[width=5in]{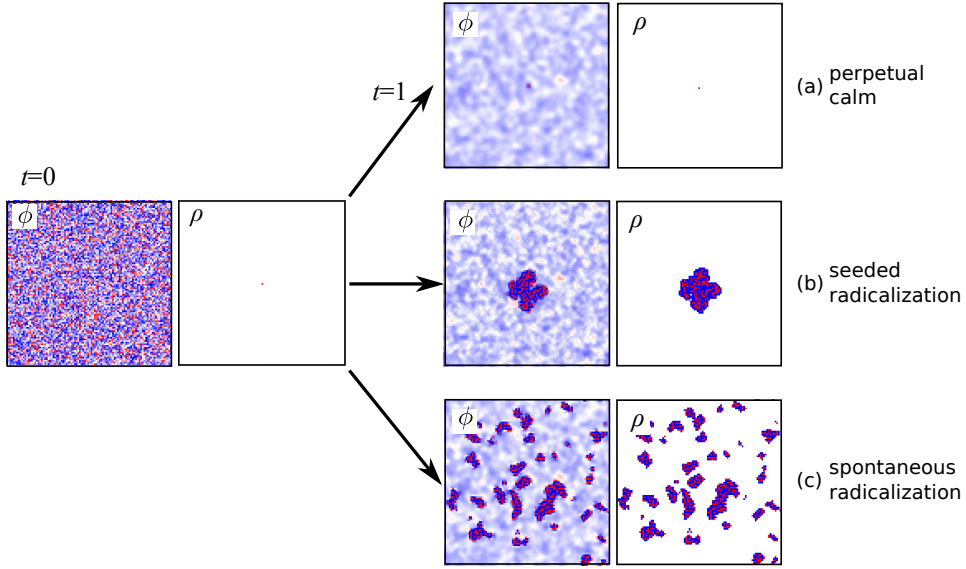}
\end{center}
  \caption{ Three typical evolutions of the model
    Eqs.~(\ref{EQ:TENSION})-(\ref{EQ:ZEAL_ADJUSTMENT}): (a) a
    perpetual calm situation ($J_-=0.01$, $J_+=0.1$), (b) seeded
    radicalization ($J_-=0.03$, $J_+=0.6$), and (c) spontaneous
    radicalization ($J_-=0.06$, $J_+=0.4$).  The left panel shows the
    initial conditions of $\phi$ and $\rho$ with an radical seed at
    the center of the lattice. In the perpetual calm situation, the
    radical seed is unable to radicalize anyone else. In the scenario
    of seeded radicalization, radical population spreads out from the
    initially seeded radical through nearest-neighbor
    interactions. For spontaneous radicalization, non-radicals turn
    radical without getting in contact with another radical agent.
    The other parameters of these simulations $D=0.1$, $R_0=3/7$, and
    $N=100$.
    \label{FIG:FATES} }
\end{figure}

This scenario can also be described as ``heterogeneous nucleation'' of
an ``antiferromagnetic'' phase within the context of solid state
physics.  In addition to nucleation by radical agents, under different
parameter regimes, our model can also exhibit other qualitatively
different dynamics, as displayed in Fig.~\ref{FIG:FATES}.  The
behaviors outlined here qualitatively represent those of all possible
parameter choices, as confirmed by extensive simulations.
%
%
Using the same initial conditions as in Fig.~\ref{FIG:COMP} but
choosing $J_{-}= 0.01, J_{+}=0.1$, Fig.~\ref{FIG:FATES}(a) depicts a
permanently calm situation where $\phi_i$ converges towards an
intermediate consensus value throughout the lattice except near the
initially planted radical.  Although its neighboring agents become
zealots, as shown by the darker blue colors, they remain non-radical
and prevent radical attitudes from spreading. We refer to this outcome
as one of ``perpetual calm.''  Fig.~\ref{FIG:FATES}(b) displays the
same results as in Fig.~\ref{FIG:COMP} for $J_{-}=0.03$ and
$J_{+}=0.6$.  We denote this behavior as ``seeded radicalization.''
Finally, in Fig.~\ref{FIG:FATES}(c), the parameters $J_{-}=0.06,
J_{+}=0.4$ lead to a hypersensitive system where non-radicals can
spontaneously radicalize.  Clusters of high tension
``antiferromagnetic'' domains spontaneously arise in a manner similar
to homogeneous nucleation. We call this type of response ``spontaneous
radicalization.'' These three scenarios comprise all qualitatively
distinct outcomes of the model seeded with a radical agent at the
center of the lattice.

To quantitatively compare the three qualitatively different outcomes
shown in Fig.~\ref{FIG:FATES}, we compute
\begin{equation}
\begin{array}{ll}
\mbox{(a) mean radicalization:} & \displaystyle \bar{\rho}(t) =
{1 \over N^{2}} \sum_{i}\rho_{i}(t)
\vspace{1mm} \\ 
\mbox{(b) ${(\phi>0)\over (\phi<0)}$ (red:blue) ratio:} & \displaystyle
R(t) = {\sum_{i} H(\phi_{i})\over \sum_{j}H(-\phi_{j})} 
\vspace{1mm} \\
\mbox{(c) mean belief value:} &  \displaystyle \bar{\phi}(t) =
{1 \over N^{2}} \sum_{i}\phi_{i}(t)
\vspace{1mm} \\ 
\mbox{(d) polarity of belief:} &  \displaystyle P(t)= {1 \over N} \sqrt{\sum_{i}(\phi_{i}(t)-\bar{\phi}(t))^{2}},
\end{array}
\end{equation}
where $H(x) = 1$ for $x>0$ and $0$ otherwise is the Heaviside
function. Here, $\bar{\phi}(t)$ can be interpreted as a consensus
belief, and $P(t)$ is the standard deviation of belief.
%
\begin{figure}[htb]
  \begin{center}
      \includegraphics[width=5in]{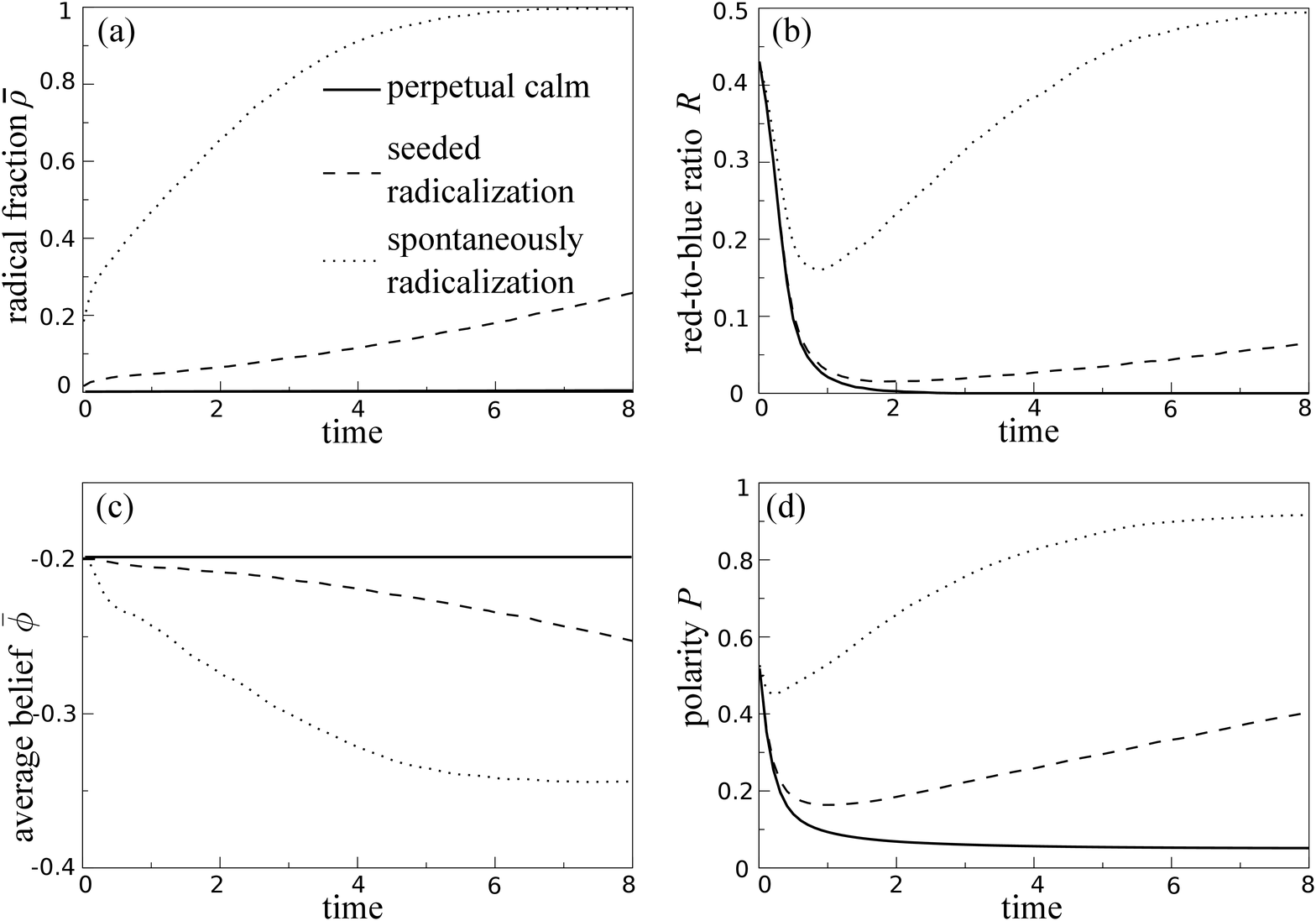}
\end{center}
  \caption{ Time series of (a) radical fraction $\bar{\rho} (t)$,
    (b) red-to-blue population ratio $R(t)$, (c) mean belief value
    $\bar{\phi} (t)$, and (d) polarity of belief $P (t)$ for
    the scenarios of perpetual calm (solid curves), seeded
    radicalization (dashed curves), and spontaneous radicalization 
    (dotted curves). For the perpetual calm situation, decreasing $P$
    indicates that individual $\phi_i$ conforms to a consensus. During
    the conformation of $\phi_i$, red population is converted to blue,
    as demonstrated by the decreasing $R(t)$. In the other two
    scenarios, the emergence of radicals eventually causes 
    $\phi_i$ to deviate from the consensus, leading to high $P$. The
    emerging radicals, mainly blue, also prompt non-radical blue
    individuals to switch sides, leading to a rising trend of
    $R(t)$. The parameters for each of the three scenarios are the
    same as in Fig.~\ref{FIG:FATES}.  
    \label{FIG:CONSENSUS} }
\end{figure}
%

Fig.~\ref{FIG:CONSENSUS}(a) shows the radical fraction $\bar{\rho}(t)$
as a function of time.  For the case of perpetual calm (solid curve),
none of the non-radical agents are turned radical by the planted
radical seed, and $\bar{\rho}(t) = 1/N^2$ throughout the simulation.
If the sensitivity $J_{\pm}$ to different neighboring beliefs is
increased, radicalization can spread radially from the radicalized
seed.  The thin dashed curve in Fig.~\ref{FIG:CONSENSUS}(a) shows that
the area fraction increases quadratically with time, implying that the
typical length scale of the ``antiferromagnetic'' radicalization phase
increases linearly in time.  If the minimum sensitivity $J_{-}$ is
further increased, tension between neighboring non-radicals with
different beliefs is not low enough to prevent spontaneous
radicalization. In this case, $\bar{\rho}(t)$ (thick curve) rises
quickly to its maximum of unity.

Fig.~\ref{FIG:CONSENSUS}(b) plots the evolution of the red-to-blue
population ratio $R(t)$. In the case of perpetual calm, belief
conformity prompts all agents to join the majority blue sect. In the
other two cases, the minority red sect members initially conform to
the blue ideology.  However, once the number of radicals increase,
some blue non-radicals become alienated by blue radicals and are
driven towards a more red belief, progressively turning into red
radicals.

Fig.~\ref{FIG:CONSENSUS}(c) shows that under calm conditions the
average opinion $\bar{\phi}(t)$ remains constant.  Under these
conditions $k (\phi_i, \phi_j, \rho_i, \rho_j)$ is symmetric for every
$i$-$j$ pair and $\dd \bar{\phi}/\dd t = 0$.  The value of
$\bar{\phi}(t) = -0.2$ is thus set by the initial red-to-blue ratio
$R(0) = 3/7$.  For the two radicalization scenarios, the increasing
number of radicals that adopt extreme beliefs causes $\bar{\phi}(t)$
to deviate from $\bar{\phi}(t=0)$.

Finally, in Fig.~\ref{FIG:CONSENSUS}(d) the polarity $P(t)$ shows
convergence of $\phi_i$ towards a consensus belief in the case of
perpetual calm.  This is typical for canonical DeGroot models, except
that the planted radical seed prevents $P(t)$ from vanishing
asymptotically.  In the case of seeded radicalization (dashed curve),
$P(t)$ initially decreases due to fast conformity followed by a slower
increase during which radicalization spreads. If radicalization is
spontaneous, the initial conformation phase of decreasing $P(t)$ is
overcome by a more rapid radicalization rate that leads to larger
polarity.

\begin{figure}[h]
  \begin{center}
      \includegraphics[width=5in]{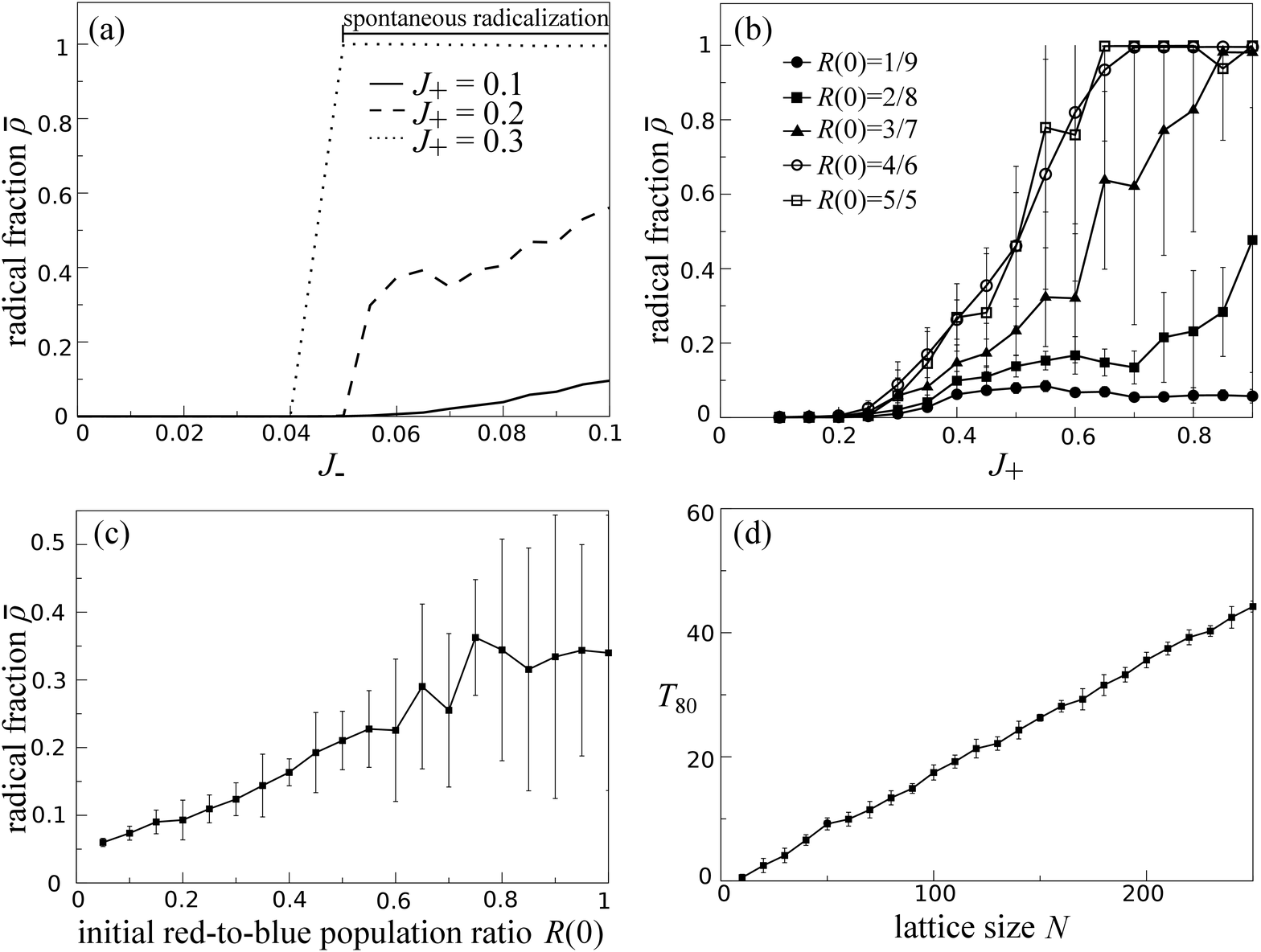}
\end{center}
  \caption{ Parameter dependence of the basic model
    (Eqs.~(\ref{EQ:TENSION})-(\ref{EQ:ZEAL_ADJUSTMENT})). (a)
    Spontaneous radicalization triggered by increasing
    $J_-$. Long-time ($t=50$) radical fractions of simulations without
    initial radical seeds are plotted against $J_-$ for various $J_+$
    values. When $J_-$ exceeds a threshold, non-radicals spontaneously
    radicalize. For long enough time and sufficiently large $J_+$, the
    spontaneously emerging radicals spread through the entire
    population. (b) Spread of seeded radicalization regulated by $J_+$.  
    Radical fractions at an intermediate time ($t=10$) are plotted
    against $J_+$ for various $R(0)$ ratios. A larger $J_+$ causes
    radicals to spread faster, reaching a higher radical fraction at
    an intermediate time. (c) Intermediate-time ($t=10$) radical
    fraction versus $R(0)$ for $R(0) \le 1$. More closely matched
    initial red and blue populations also result in faster spread of
    radicals. (d) Time for the radicalization cluster to reach $80\%$ 
    of the system area for various lattice domain size $N$. The time
    increases linearly with $N$, suggesting a linear radial expansion
    of the cluster over time. If not varied in the figures, the
    default parameters values are used. An initial radical seed is planted for
    figures (b)-(d) but not (a). Each data point represents the mean value of
    ten simulations, and error bars the standard deviations.
    \label{FIG:J_MINUS}}
\end{figure}

We find that the sensitivity of non-radicals $J_-$ is the primary
determinant of whether spontaneous radicalization emerges or not.  In
Fig.~\ref{FIG:J_MINUS}(a), we plot radical fractions versus $J_-$ for
several values of $J_+$ at a long time after initiation ($t=50$) to
identify the parameter regimes where spontaneous radicalization
arises.  Initial conditions are set at $\sigma_i = 0$ and a randomly
distributed $-1 \le \phi_i \le 1$ with $R(0) = 3/7$.  No radical
agents are planted at $t=0$.
%
%
In the absence of a radical seed, non-radicals become radicalized
exclusively through the tensions arising from belief differences
amongst themselves.  We find that spontaneous radicalization is
triggered for $J_- > 0.04$ and that this threshold does not depend on
$J_+$.  For low values of $J_+ \lesssim 0.2$, the spread of radicals
can be arrested after the emergence of spontaneously radicalized
patches.  As a result, radicals do not pervade society even at long
times.

Henceforth, we plant a radical seed at the center and set $J_- =0.03$
to focus on seeded radicalization, a qualitatively reasonable
description of the nucleation and growth of sectarianism.  Recalling
that $\sigma_i \le 32 {\rm max}(J)$, we have $\sigma_i \le 0.96 < 1$
if $J=J_-$ everywhere, eliminating the chance of spontaneous
radicalization. As long as $J_- < 1/32$, spontaneous radicalization
cannot arise.  In Fig.~\ref{FIG:J_MINUS}(b), we plot radical fractions
$\bar{\rho}(t=10)$ as a function of $J_{+}$ for various initial
red-to-blue ratios $R(0)$. Radicals begin to spread from the planted
seed when $J_+ > 0.25$ regardless of $R(0)$. For larger $J_{+}$, the
radicalization cluster reaches a larger fraction of the lattice
indicating a faster spreading rate. A larger initial ratio $R(0)$ also
causes the cluster of radicals to spread at a faster rate.  This is
confirmed in Fig.~\ref{FIG:J_MINUS}(c) where radical fraction
$\bar{\rho}(t=10)$ for $J_{+}=0.6$ increases with $R(0)$, and is
maximal for $R(0) \gtrsim 0.75$, where the members of the two sects
are about equal.

These findings are consistent with the observation that conflicts
mostly arise in regions where ethnic boundaries were not well-defined
(\textit{i.e.}, a mixed population) and where the populations of
ethnic groups are closely matched \cite{LIM07}.  A minority population
that is overwhelmed by a much more populous opposing belief more
easily assimilates and is less likely to elicit conflict. An example
can be found in Indonesia, where analysis of survey data suggested
that the spread of radical beliefs was subdued in villages consisting
of a notably dominant majority population \cite{BAR04}.

In Fig.~\ref{FIG:J_MINUS}(d) we plot the time $T_{80}$ for the
radicalization cluster to cover $80\%$ of the lattice, which for our
simulations corresponds roughly to the time it takes for the cluster
perimeter to reach the boundaries of the lattice.  We find that
$T_{80}$ increases linearly with domain size $N$, suggesting that the
radius of radicalized cluster area grows linearly with time, and that
the corresponding area scales as $t^{2}$, consistent with
Fig.~\ref{FIG:CONSENSUS}(a).

Finally, we find that tension $\sigma_i$ rarely decreases among a
mixed population, precluding de-radicalization. As a consequence, we
find that the value of $D$ has essentially no effect on seeded
radicalization in our basic model
Eqs.~(\ref{EQ:TENSION})-(\ref{EQ:ZEAL_ADJUSTMENT}).

\subsection{Global institutional influence}

So far we have investigated the parameter dependence of the basic
model Eqs.~(\ref{EQ:TENSION})-(\ref{EQ:ZEAL_ADJUSTMENT}).  We now
explore how an external institutional influence may affect
radicalization.  We set the basic model parameters to the default
values ($J_-=0.03$, $J_+=0.6$, $D=0.1$, $R(0)=3/7$, and $N=100$) and
focus on the fast seeded radicalization regime, shown in
Fig.~\ref{FIG:FATES}(b) in the absence of any external players. We
here add a global institutional influence $G(\phi_i; \phi_{0})$ by
replacing Eq.~(\ref{EQ:TENSION}) with Eq.~(\ref{EQ:TENSION2}).

\begin{figure}[h]
  \begin{center}
      \includegraphics[width=5in]{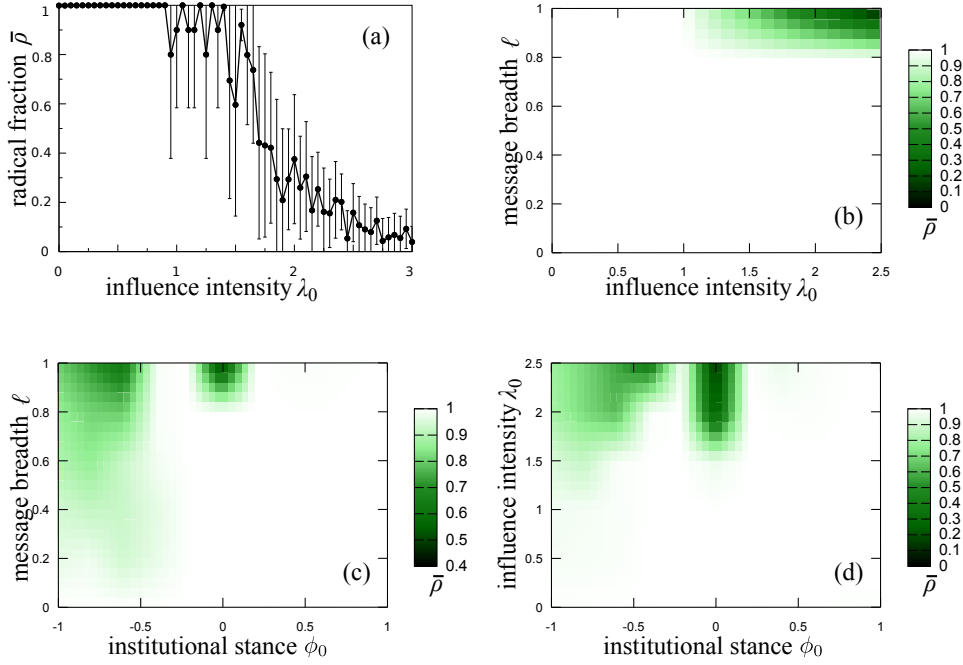}
\end{center}
  \caption{ (a) Reduction of radical fraction for various strengths of
    global institutional influence $\lambda_0$. We plot long-time
    ($t=50$) radical fractions versus $\lambda_0$ while setting
    institutional stance $\phi_0 = 0$ and message breadth $\ell =
    1$. When $\lambda_0 \gtrsim 1.6$, the institutional influence
    begins to show significant effect, and radicals have mostly
    stopped spreading when $\lambda_0 \gtrsim 2.5$. (b) Long-time
    ($t=50$) radical fractions versus $\ell$-$\lambda_0$, (c)
    $\ell$-$\phi_0$, and (d) $\lambda_0$-$\phi_0$. The color maps
    represent radical fractions. For (b) the institution adopts a
    neutral stance $\phi_0=0$. The most reduction of radicals is
    achieved at high $\lambda_0$ and large $\ell$. For (c), we set
    $\lambda_0 = 2$. A neutral ($\phi_0$) and a majority-biased stance
    ($\phi_0 \sim -0.5$) both register significant reduction of
    radicals. For (d) we choose $\ell = 1$, and again $\phi_0=0$ and
    $\phi_0 \sim 0.5$ results in significant reduction of
    radicals. Each data point represents the mean value of ten
    simulations, and error bars in (a) the standard deviations. 
  \label{FIG:GLOBAL} }
\end{figure}

In Fig.~\ref{FIG:GLOBAL}(a) we plot the long-time ($t=50$) radical
fraction $\bar{\rho}(t=50)$ and examine the effect of $\lambda_0$,
which defines the intensity of the tension-reducing influence.  We
choose $\phi_0 = 0$ and $\ell = 1$, so that the external institution
adopts a neutral stance and reduces perceived tension for individuals
with any belief value $\phi_i$.  For these parameters, we observe
significant and consistent reduction of radical fractions when
$\lambda_0 \gtrsim 1.6$.  For $\lambda_0 \gtrsim 2.5$, the spread of
radicals by the seed is largely suppressed.  Hence, one of our major
findings is that to exert significant influence $\lambda_{0}$, the
institutional influence needs to have a high penetration within the
overall population.
 
In Fig.~\ref{FIG:GLOBAL}(b), the radical fraction $\bar{\rho}(t = 50)$
is depicted using a color intensity map and plotted as a function of
$\ell$ and $\lambda_0$. As expected, the lowest radicalization levels
are achieved by large $\lambda_0$ and $\ell$, indicating that for a
strong influence intensity to pacify conflicts, the institutional
publicity must also have broad appeal.  Note that in realistic
situations the institutional influence intensity and message breadth
are often constrained by the resources available to the
institution. It may thus become impractical to simultaneously achieve
high penetration and broad appeal given limited resources. How to most
effectively allocate resources is an interesting optimization problem.

%
%

In Fig.~\ref{FIG:GLOBAL}(c) we plot $\bar{\rho}(t = 50)$ against
$\ell$ and $\phi_0$ with $\lambda_0 = 2$. The largest reduction of
radicals is occurs in the tongue near the neutral institutional stance
$\phi_0 = 0$, but diminishes as $\ell$ is decreased.  Some reduction
of radicals is also observed when $-1< \phi_0 < -0.5$, corresponding
to a stance biased toward the majority belief. Although this latter
regime $-1< \phi_0 < -0.5$ does not result in as significant a
reduction in radical level as the $\phi_0\approx 0$ tongue for $\ell
\approx 1$, the reduction occurs over a wider range of $\ell$.  We
thus find that if the institution is unable to fashion a message with
sufficiently broad appeal, it may be better bias the influence to
appease the majority.

In Fig.~\ref{FIG:GLOBAL}(d) we show $\bar{\rho}(t = 50)$ versus
$\lambda_0$ and $\phi_0$ with $\ell = 1$. Again, a neutral
institutional stance ($\phi_0=0$) achieves the most reduction of
radicals, while a majority-biased stance also has some success but to
a lesser degree. With respect to influence intensity $\lambda_0$, we
find that the effectiveness of a majority-biased diminishes quickly
with decreasing $\lambda_0$, while a neutral stance is capable of
maintaining a better result at a lower $\lambda_0$.

Our results suggest that an institutional influence achieves optimal
results if the entity carefully adopts a strong but neutral stance
between the two conflicting beliefs. However, the outreach of the
institutional message content is also important. If the institutional
influence targets a narrow range of beliefs for the reduction of
perceived tension, it may alienate those out of reach, and may have
the opposite effect of increasing radicalization.  If the institution
is unable to placate population with a wide range of beliefs, a stance
favoring the majority view may be an effective alternative. Of course,
other mechanisms of external influence may apply. For example,
governing institutions may directly influence people's beliefs rather
than just the tension they perceive.  To model such mechanisms, a
modification of Eq.~\ref{EQ:ZEAL_ADJUSTMENT} can be developed.

\section{Conclusions}

In this paper, we construct a belief dynamics model that incorporates
a bistable radicalization process to describe the spread of sectarian
conflict.  While the model equations can be applied to arbitrary
social network structures, we simulate our model on a locally
connected two-dimensional square lattice with periodic boundary
conditions. By defining belief and radicalization as separate
variables, our model allows for a more nuanced description that
distinguishes belief from radicalization. Although in the absence of
radicals, all non-radicals asymptotically conform to an
apathetic/neutral consensus due to the simplicity of the model, we do
observe transient enclaves of more zealous but non-radical
individuals. Radicals, on the other hand, are mostly zealots with
extreme belief values. We examine the parameter dependence of the
model and identify regimes leading to three distinct evolution
paths. In the regime of perpetual calm, non-radical individuals cannot
be radicalized even if by planting radical seeds in advance. In the
regime of spontaneous radicalization, non-radical individuals may
spontaneously radicalize even in the absence of radicals. Between the
above two regimes lies the regime of seeded radicalization, where
non-radical individuals cannot spontaneously radicalize but can become
radical upon contact with other radicals. For subsequent
investigations, we choose parameter values in the third regime, as the
most realistic scenario for the propagation of sectarian conflicts.
We find that radicalization can be suppressed by a numerically more
dominant majority population between the two competing sects.  Finally
we implement institutional influence as a globally connected player
and find that the most effective intervention to pacify conflict is to
adopt a strong but neutral view.

Our model represents a first step in studying the bistable (or
multistable) nature of human behaviors in the development of social
conflicts. Our incorporation of bistability is only phenomenological,
while the underlying mechanism is a fundamental but much more
challenging to design and include.  Moreover, the separation of
ideological belief and radicalization that we propose is a simplified
version of multi-dimensional opinion dynamics models. Our modeling
framework can be extended to include multiple ideological spectrums,
such as religion, social economics, and politics, and examine the
interplay among them.  
Finally, our model can be straightforwardly extended to include
evolving, non-lattice social networks.

\bibliographystyle{amsplain}
\bibliography{crime}

\appendix

\renewcommand{\theequation}{A\arabic{equation}}

\section*{Appendix: Numerical implementation}

For numerical simulations, we adopt a semi-implicit method with a
fixed time step size to integrate our model.  Let us denote $\phi_i
(t)$, $\rho_i (t)$, and $\sigma_i(t)$ at a discrete time $t = n \Delta t$
as $\phi_i^n$, $\rho_i^n$, and $\sigma_i^n$, where $\Delta t$ is the time
step size. Then we discretize
Eqs.~(\ref{EQ:TENSION})-(\ref{EQ:ZEAL_ADJUSTMENT}) as

\begin{eqnarray}
    \sigma_i^{n+1} & = & \sum_{j \in {\rm n.n.[i]}}
       J(\rho_i^{n+1}, \rho_j^{n+1}) \left( \phi_i^{n+1} - \phi_j^{n+1} \right)^2,
   \label{EQ:DISCRETE_TENSION} \\
      \rho_i^{n+1} & = & \left\{ \begin{array}{ll}
         1 & \textrm{if } \sigma_i^{n+1} > 1, \\
         0 & \textrm{if } \sigma_i^{n+1} < D, \\
         \rho_i^n & \textrm{otherwise},
   \end{array}\right.
   \label{EQ:DISCRETE_RHO} \\
    \phi_i^{n+1} & = & \phi_i^n
                    + \Delta t \sum_{j \in {\rm n.n.[i]}} k(\rho_i^n, \rho_j^n, \phi_i^n, \phi_j^n) 
                           \left( \phi_j^n - \phi_i^n \right), 
    \label{EQ:DISCRETE_ZEAL_ADJUSTMENT} 
\end{eqnarray} 
and an iterative method is used to solve the semi-implicit equations
(\ref{EQ:DISCRETE_TENSION})-(\ref{EQ:DISCRETE_ZEAL_ADJUSTMENT}). The
equations with global institutional influence are solved in the same way.
Note that for an explicit method, Eq.~(\ref{EQ:RHO}) may impose a
severe constraint on $\Delta t$, and 
even with an adaptive time step size, the numerical integration can
still be very inefficient. 

\end{document}